\documentclass[aps,prl,twocolumn,amsfonts,amssymb,amsmath,floatfix,10pt,superscriptaddress]{revtex4-1}
\usepackage[latin1]{inputenc}
\usepackage[T1]{fontenc}
\usepackage[english]{babel}
\usepackage{graphicx}
\usepackage{ulem}
\usepackage{hyperref}
\usepackage{siunitx}

\DeclareSIUnit{\Erecoil}{E_r}
\DeclareSIUnit{\Tfermi}{T_F}
\DeclareSIUnit{\kbrillouin}{k_{BZ}}
\DeclareSIUnit{\ab}{a_{0}}
\DeclareSIUnit{\abb}{a_{bb}}
\DeclareSIUnit{\abf}{a_{bf}}

\begin{document}

\title{Experimental reconstruction of the Berry curvature in a topological Bloch band}

\author{N.~Fl\"aschner}
\altaffiliation{These authors have contributed equally to this work}.
\affiliation{Institut f\"ur Laser-Physik, Universit\"at Hamburg, Luruper Chaussee 149, 22761 Hamburg, Germany}
\affiliation{The Hamburg Centre for Ultrafast Imaging, Luruper Chaussee 149, 22761 Hamburg, Germany}

\author{B. S.~Rem}
\altaffiliation{These authors have contributed equally to this work}.
\affiliation{Institut f\"ur Laser-Physik, Universit\"at Hamburg, Luruper Chaussee 149, 22761 Hamburg, Germany}
\affiliation{The Hamburg Centre for Ultrafast Imaging, Luruper Chaussee 149, 22761 Hamburg, Germany}

\author{M.~Tarnowski}
\affiliation{Institut f\"ur Laser-Physik, Universit\"at Hamburg, Luruper Chaussee 149, 22761 Hamburg, Germany}

\author{D.~Vogel}
\affiliation{Institut f\"ur Laser-Physik, Universit\"at Hamburg, Luruper Chaussee 149, 22761 Hamburg, Germany}

\author{D.-S.~L\"uhmann}
\affiliation{Institut f\"ur Laser-Physik, Universit\"at Hamburg, Luruper Chaussee 149, 22761 Hamburg, Germany}

\author{K.~Sengstock}
\altaffiliation{sengstock@physnet.uni-hamburg.de}
\affiliation{Institut f\"ur Laser-Physik, Universit\"at Hamburg, Luruper Chaussee 149, 22761 Hamburg, Germany}
\affiliation{The Hamburg Centre for Ultrafast Imaging, Luruper Chaussee 149, 22761 Hamburg, Germany}
\affiliation{ZOQ - Zentrum f\"ur Optische Quantentechnologien, Universit\"at Hamburg, Luruper Chaussee 149, 22761 Hamburg, Germany}

\author{C.~Weitenberg}
\affiliation{Institut f\"ur Laser-Physik, Universit\"at Hamburg, Luruper Chaussee 149, 22761 Hamburg, Germany}
\affiliation{The Hamburg Centre for Ultrafast Imaging, Luruper Chaussee 149, 22761 Hamburg, Germany}

\date{\today}



\maketitle

\textbf{Topological properties lie at the heart of many fascinating phenomena in solid state systems such as quantum Hall systems or Chern insulators. The topology can be captured by the distribution of Berry curvature, which describes the geometry of the eigenstates across the Brillouin zone. Employing fermionic ultracold atoms in a hexagonal optical lattice, we generate topological bands using resonant driving and show a full momentum-resolved measurement of the ensuing Berry curvature. Our results pave the way to explore intriguing phases of matter with interactions in topological band structures. }

\begin{figure}[h]
  \centering
  \includegraphics[width=\columnwidth]{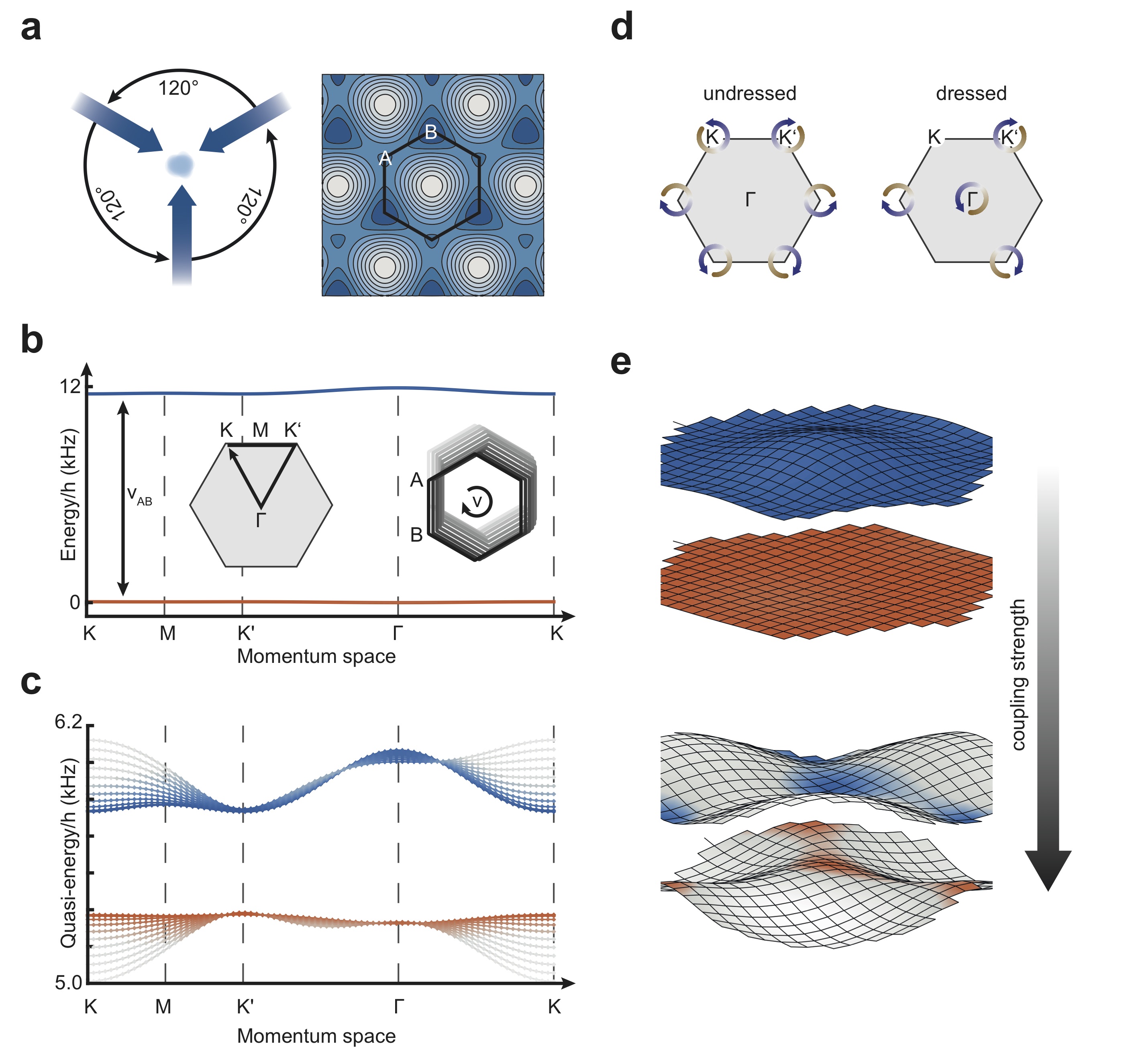}
  \caption{\textbf{Engineering of the topology by band dressing.} (a) Three laser beams intersecting under 120$^\circ$ interfere to form a tunable honeycomb lattice~\cite{SupplMat} with a variable offset energy $h \Delta_{AB}$ between A and B sites. (b) Circular shaking of the lattice (shaded inset on the right) with frequency $\nu$ resonantly ($\nu \approx \nu_{AB}$) couples the two lowest bands, which are plotted (red and blue lines) along a high symmetry path (K,M,K',$\Gamma$, K). (c) Dressed Floquet bands for different shaking amplitudes between $0$ and $223$~nm at a shaking frequency of $\nu = 11$~kHz. The dressed bands are calculated using Floquet theory~\cite{SupplMat}. (d) Sketch of the position of the topological defects in the undressed and dressed cases, illustrating the dramatic change of the topology. The arrow indicates the direction of the phase winding around the Dirac point. (e) 2D dispersion relations of the bare and dressed bands showing a six-, respectively three-fold symmetry. In (c and e), the color represents the dressing (less color corresponds to stronger dressing).}
  \label{fig1}
\end{figure}

Topology is a fundamental concept for our understanding of many fascinating systems that have recently attracted a lot of interest, such as topological superconductors or topological insulators, which conduct only at their edges~\cite{Hasan2010}. The topology of the bulk band is quantified by the Berry curvature~\cite{Xiao2010} and the integral over the full Brillouin zone is a topological invariant, called the Chern number. According to the bulk boundary correspondence principle, the Chern number determines the number of chiral conducting edge states~\cite{Hasan2010}. While in a variety of lattice systems ranging from solid state systems to photonic waveguides and even coupled mechanical pendula, edge states have been directly observed~\cite{Hafezi2013,Rechtsman2013,Wang2009,Ningzuan2015,Susstrunk2015}, the underlying Berry curvature as the central measure of topology is not easily accessible. In recent years, ultracold atoms in optical lattices have emerged as a platform to study topological band structures~\cite{Goldman2014,Kitagawa2010} and these systems have seen considerable experimental and theoretical progress. Whereas in condensed matter systems, topological properties arise due to external magnetic fields or intrinsic spin-orbit coupling of the material, they can in cold atom systems be engineered by periodic driving analogous to illuminated graphene~\cite{Sie2014}. Interestingly, the resulting Floquet system can have totally new topological properties~\cite{Bukov2015}. The driving can, for example, be realized by lattice shaking~\cite{Lignier2007,Struck2012,Struck2013,Parker2013,Jotzu2014} or Raman coupling~\cite{Lin2011,Aidelburger2015,Kennedy2015} with high precision control in a large parameter space. In particular, the driving can break time-reversal symmetry~\cite{Struck2012,Struck2013,Jotzu2014} and thus allows for engineering non-trivial topology~\cite{Jotzu2014,Aidelburger2015}. In quantum gas experiments, topological properties have been probed via the Hall drift of accelerated wave packets~\cite{Jotzu2014,Aidelburger2015}, via an interferometer in momentum space~\cite{Duca2015,Li2015} and in charge pump experiments~\cite{Lu2015} but so far, the underlying Berry curvature was not revealed. 

Here, we demonstrate a method~\cite{Hauke2014,Alba2011} that, for the first time, allows for a direct measurement of the Berry curvature with full momentum resolution. We perform a full tomography of the Bloch states across the entire Brillouin zone by observing the dynamics at each momentum point after a projection onto flat bands. The topological bands are engineered by resonant dressing of the two lowest bands of an artificial boron nitride lattice and feature a rich distribution of Berry curvature. From the Berry curvature, being the fundamental building block of topology, other relevant quantities such as the Berry phase or the Chern number can easily be obtained.

\begin{figure}[b]
  \centering
  \includegraphics[width=\columnwidth]{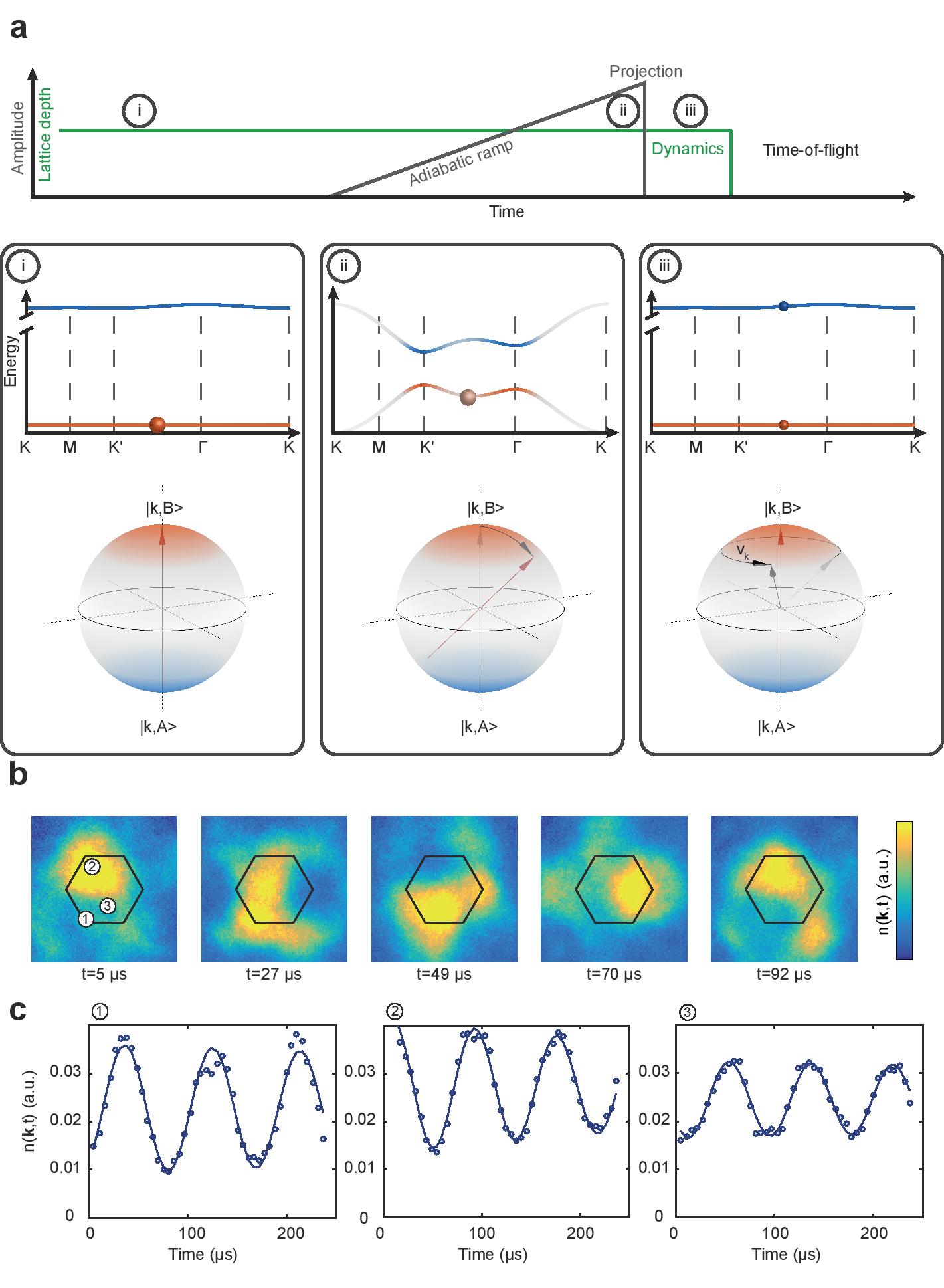}
  \caption{Dynamical measurement of the topology of the dressed bands. (a) Illustration of the experimental protocol.  We start in a band insulator in the lowest undressed band (i) and adiabatically ramp up the coupling strength (shaking amplitude) (ii). Now the Bloch state at each quasimomentum is a superposition of $\lvert\mathbf{k},A\rangle$ and $\lvert\mathbf{k},B\rangle$, pointing in different directions on the Bloch sphere. When the dressing is switched off (iii), the state is projected onto the flat bands and rotates on the Bloch sphere with the band difference $\nu_{\mathbf{k}}$. After time-of-flight, this yields an oscillation of the density at each momentum. (b) Experimental momentum distributions for different hold times after the projection onto flat bands. The color is the atomic density in momentum space (after time-of-flight), the hexagon marks the first Brillouin zone. Note that the clearly visible oscillations of the density at different momenta stem from interference of the two bands (Eq. (2)). (c) Oscillations at different quasimomenta. The solid lines are sinusoidal fits. From the amplitude and phase of the oscillation, we reconstruct the dressed state according to Eq. (1). The experimental parameters are $V_l = 15.15(15)~E_r$, $\Delta_{AB} = 11.45(11)$~kHz, $\nu = 11$~kHz and a shaking amplitude of $223$~nm~\cite{SupplMat}.
}
  \label{fig2}
\end{figure}

Our system consists of ultracold fermionic atoms in a hexagonal optical lattice~\cite{SoltanPanahi2011} formed by three interfering laser beams. With an appropriate polarization~\cite{SupplMat}, a variable energy offset $h\Delta_{AB}$ between the A and B-sites (Fig. 1a), which breaks inversion symmetry, can be engineered. With the emerging band gap $h \nu_{AB}$, the Dirac points at K and K' become massive and for a large offset, the bands are flat (Fig. 1b)~\cite{SupplMat}. This is a key ingredient for our tomography, since the flat band acts as the reference frame in which we reconstruct the eigenstates. Now, as a central experimental feature, we can dress the bands via circular shaking. In Fig. 1c, we show the dressed Floquet bands for near resonant circular shaking for different accessible driving amplitudes. Apart from the dramatic change in the dispersion relation, the topological properties of the bands are changed. This manifests itself in the creation of a new Dirac point at the $\Gamma$-point and the annihilation of a Dirac point at the K point (Fig. 1d). A three-fold symmetry also becomes visible in the dispersion relation (Fig. 1e). 

\begin{figure}[h!]
  \centering
  \includegraphics[width=\columnwidth]{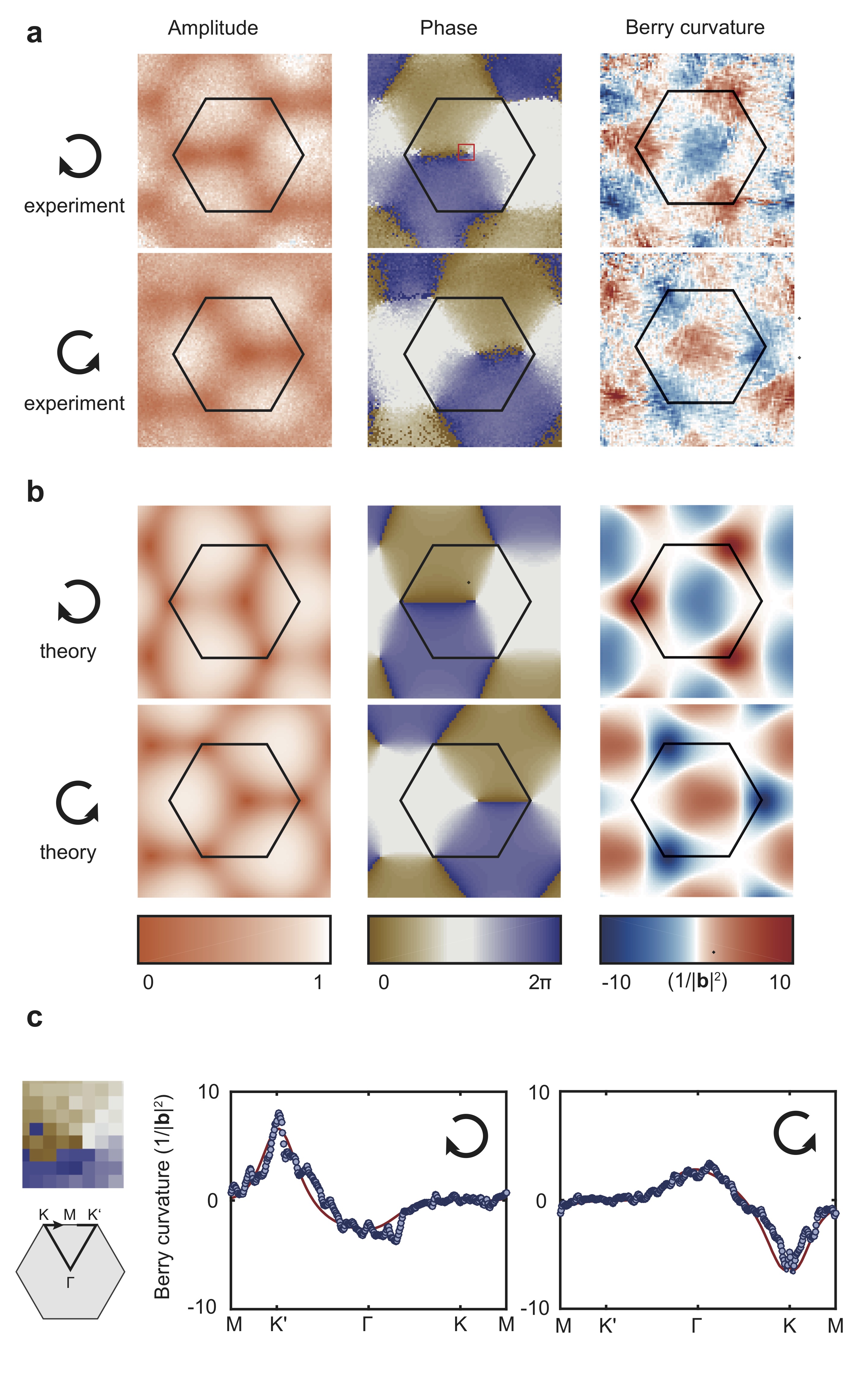}
  \caption{Momentum-resolved measurement of the Berry curvature. (a) Amplitude (left column) and phase (central column) obtained from the fits to the oscillations from Fig. 2, performed for each pixel in and around the first Brillouin zone (hexagon). From those fit results, we obtain, as our central result, the momentum-resolved Berry curvature (right column), (see text for details). The experiment was performed for different chiralities of the lattice shaking, destroying the Dirac point at either K or K'. (b) Theoretical results from a tight-binding Floquet calculation~\cite{SupplMat} using precisely the experimental parameters, yielding a very good agreement. (c) In the upper left corner, a zoom into a phase vortex (red square in (a)) is shown, illustrating the high momentum resolution of our method due to a pixel-wise evaluation of more than 2800 pixels in the first Brillouin zone. The plots show the experimental Berry curvature (blue points) along the high-symmetry path in comparison with the theoretical (red solid lines) calculation. In (b) and (c), the Berry curvature is given in units of the inverse reciprocal lattice vector length $\lvert \mathbf{b} \lvert$ squared~\cite{SupplMat}.}
  \label{fig3}
\end{figure}

The topological properties are not captured by the mere dispersion relation but by the Berry curvature, which describes the winding of the eigenstates across the Brillouin zone. Therefore, a complete tomography of the eigenstates of a Bloch band is mandatory for a measurement of the Berry curvature. The key idea behind our tomography is to reconstruct the eigenvectors from dynamics after a projection onto flat bands~\cite{Hauke2014}. Consider the Bloch sphere (Fig. 2a) whose poles are given by $\lvert\mathbf{k},A\rangle$ and $\lvert\mathbf{k},B\rangle$, which are the Bloch states restricted to the A- respectively B-sublattice. The lower band can be written as $\lvert\mathbf{k}\rangle = \sin (\theta_{\mathbf{k}}/2) \lvert\mathbf{k},A\rangle - \cos (\theta_{\mathbf{k}}/2) \exp (i \phi_{\mathbf{k}}) \lvert\mathbf{k},B\rangle$ and after a projection onto flat bands, the state oscillates around $\lvert\mathbf{k},B\rangle$ with the frequency $\nu_{\mathbf{k}}$ given by the energy difference of the flat bands. Then, the momentum distribution after time-of-flight is given by
\begin{align}
	n(\mathbf{k}, t) &= f(\mathbf{k})\left[ 1 - \sin (\theta_{\mathbf{k}}) \cos (\phi_{\mathbf{k}} + 2 \pi \nu_{\mathbf{k}} t) \right]
	\label{eq1}
\end{align}
from which both $\theta_{\mathbf{k}}$ and $\phi_{\mathbf{k}}$ can be easily obtained, yielding the desired tomography of the eigenstates for each quasimomentum. Our method hence allows for a direct reconstruction of the Berry curvature according to 
\begin{align}
	\Omega(\mathbf{k}) &= \frac{1}{2} \left( \partial_{k,x} \mathbf{\hat{h}} \times \partial_{k,y} \mathbf{\hat{h}} \right) \cdot \mathbf{\hat{h}}
	\label{eq2}
\end{align}
with $\mathbf{\hat{h}} = (\sin (\theta_{\mathbf{k}}) \cos (\phi_{\mathbf{k}}), \sin (\theta_{\mathbf{k}}) \sin (\phi_{\mathbf{k}}), \cos (\phi_{\mathbf{k}}))$~\cite{Hauke2014}. Our experimental sequence for this state tomography is sketched in Fig. 2. We start with a cloud of $5\cdot10^4$ single-component fermionic $^{40}$K atoms forming a non-interacting band insulator in the undressed lattice. Due to a large offset between the A and B sites, $\Delta_{AB} = 11.45(11)$~kHz , the undressed bands are flat, such that for all quasimomenta $\lvert\mathbf{k}\rangle \approx \lvert\mathbf{k},B\rangle$. We adiabatically ramp up the shaking amplitude to $223$~nm within $5$~ms at a shaking frequency of $9$~kHz and then ramp the frequency to $11$~kHz within $2$~ms. By suddenly switching off the dressing, we project onto the bare flat bands, such that Eq.~\eqref{eq1} can be applied.  In Fig. 2b, we show typical time-of-flight images for different hold times in the flat bands. The images feature dynamics with very large contrast. The insets show time evolutions for different quasimomenta, revealing the pure sinusoidal oscillations with clearly distinct amplitudes $\sin (\theta_{\mathbf{k}})$ and phases $\phi_{\mathbf{k}}$ which are obtained by a simple fit to Eq.~\eqref{eq1}. We observe very large and long-lived oscillations after the projection, yielding relative amplitudes of up to 0.8. Additionally, with more than 2800 pixels in the first Brillouin zone, the resolution in momentum space is very high.

As the central result, we reconstruct the Berry curvature of the dressed band structure from these fits, which fully visualize the Bloch states as shown in Fig. 3a. The amplitude map features a pronounced three-fold symmetry, illustrating the breaking of equivalence between the K and K' points. The amplitude has a maximum at the K point and is zero at the K' and $\Gamma$ points. Even more striking is the very distinct threefold symmetry of the phase map with nearly discrete values of $\pi/3$, $\pi$, $5\pi/3$. Where the amplitudes are zero at the K' and $\Gamma$ points, the phase map correspondingly displays vortices. The phase vortices are clear signatures of Dirac points which constitute topological defects. Furthermore, the data clearly shows that we annihilate the Dirac point of the undressed hexagonal lattice at K and created a Dirac point in the dressed system at the $\Gamma$ point, thereby strikingly changing the topology of the band. The resulting Berry curvature is localized at the new Dirac points and also shows this clear three-fold symmetry. It has opposite sign at the two Dirac points which results from the opposite chirality of the phase vortices. By inverting the chirality of the shaking, we instead annihilate the Dirac point at the K' point and invert the chirality of the phase windings, which also results in an inverted symmetry in the Berry curvature. All quantities agree well with a Floquet theory calculation (see Fig. 3b), based on a tight-binding model as described in~\cite{SupplMat}. In Fig. 3c, we plot the Berry curvature pixel-wise evaluated along a high-symmetry path, illustrating the very good agreement with the theory.

As mentioned before, with the fully momentum-resolved Berry curvature, we can easily obtain the further relevant quantities like the Berry phase or Chern number. For a discussion of the respective Berry phases, see~\cite{SupplMat}. The integral over the closed area of the full first Brillouin zone must be quantized to $2\pi$ times the integer Chern number $C$. From our data we obtain $C = 0.010(13)$ and $C = -0.019(13)$ for the two different shaking chiralities~\cite{SupplMat}, which clearly confirms this quantization within the experimental errors. Our data constitutes the first momentum-resolved measurement of the Berry curvature of a topological band structure, thereby providing a deep insight into the fascinating physics of topological systems. Our measurements demonstrate that even when the global topology has Chern number zero, the distribution of Berry curvature can be very rich.

Our measurement scheme can be readily extended to characterize bands with Chern numbers different from zero~\cite{Jotzu2014,Aidelburger2015}. In principle, one could start in a shallow lattice, where reaching non-zero Chern numbers is possible, and for the tomography project onto flat bands, which can be reached, e.g., by dynamical control over the offset. Our method for generating the topological bands is spin-independent and does not couple different spin states. It therefore can be extended to high-spin systems~\cite{Krauser2014} or to strongly-interacting spin mixtures, which are expected to lead to interesting many-body phases~\cite{Neupert2011,Grushin2014,Cooper2013}.

We acknowledge stimulating discussions with Andr\'e Eckardt, Christoph Str\"ater and Maciej Lewenstein. This work has been supported by the excellence cluster "The Hamburg Centre for Ultrafast Imaging - Structure, Dynamics and Control of Matter at the Atomic Scale" and the GrK 1355 of the Deutsche Forschungsgemeinschaft.

\cleardoublepage

\onecolumngrid
\begin{center}
\textbf{\large Supplemental Material - Experimental reconstruction of the Berry curvature in a topological Bloch band}
\end{center}
\appendix
\twocolumngrid

\section{\label{secS1}A Tunable driven hexagonal lattice}
\begin{figure}[h]
  \centering
  \includegraphics[width=\columnwidth]{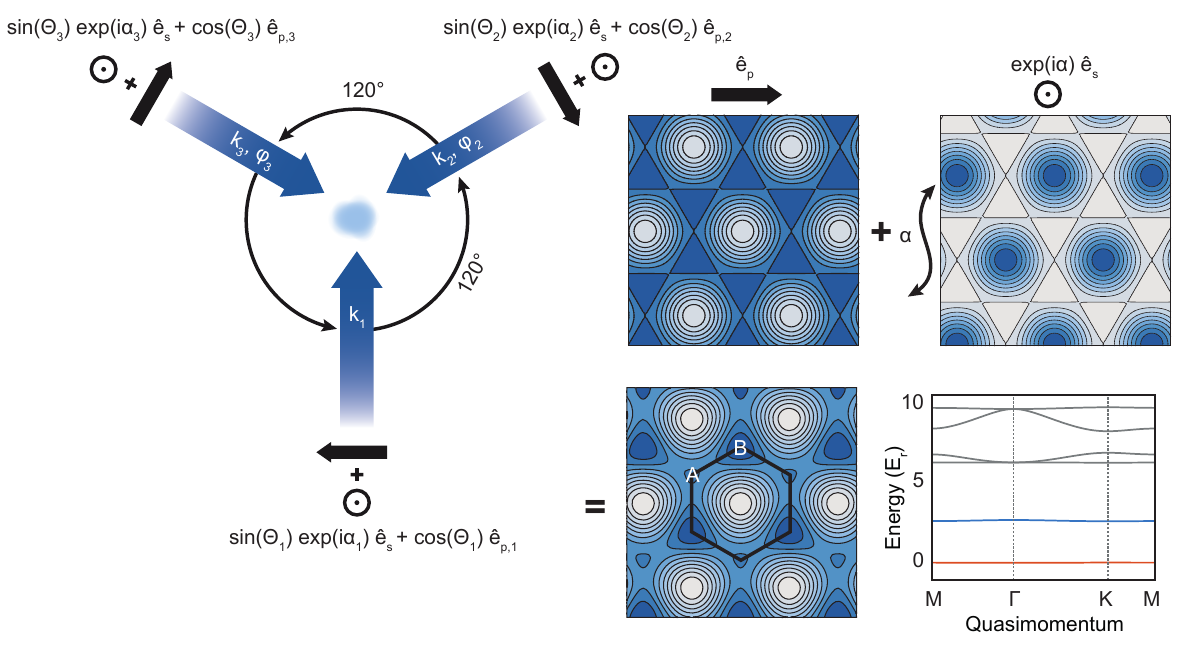}
  \caption{\textbf{A tunable hexagonal lattice setup.} The lattice is formed by three interfering laser beams intersecting under an angle of $120^\circ$. The periodic driving is realized by modulating the relative phases $\varphi$ of the beam. The p- (in-plane) polarization interference pattern is a honeycomb lattice while the s- (out-of-plane) polarization interference pattern is inverted and forms a triangular lattice. The relative phase $\alpha$ between s- and p-polarization shifts the triangular lattice with respect to the honeycomb lattice. The population $\theta$ determines the relative depth of the two lattices. For $\theta = \pi/20$ for all beams and $\alpha_{1,2,3} = 1/3 ( 0, 2\pi, 4\pi )$, a boron-nitride lattice with distinct A and B sites is realized. The resulting dispersion relation is plotted for the experimental parameters as in Fig. 2 of the main text, showing that the lower two bands (red, blue) are well separated from the other bands (gray).}
  \label{figS1}
\end{figure}

For the experiments presented in the main text, we employ a new lattice setup allowing for a variety of lattice geometries by controlling the polarizations of the interfering laser beams, following a proposal as described in~\cite{Baur2014}. The lattice potential is given by the sum of a honeycomb lattice $V_p(\mathbf{r})$ generated by the p-polarization components of the laser beams and a triangular lattice $V_s(\mathbf{r})$ due to the s-polarization components. By tuning the populations $\theta$ and relative phases $\alpha$ between the s- and p-components, the relative depth and position of the triangular lattice with respect to the honeycomb lattice can be set. In particular, we can continuously transform our pure and totally symmetric honeycomb lattice into a boron-nitride lattice in which the A- and B-sites are no longer energetically degenerate and inversion symmetry is broken (see Fig. 1a in the main text and Fig.~\ref{figS1}), as done for the experiments presented in the main text. The periodic driving of the lattice is realized by modulating the frequencies of beams 2 and 3 with acousto-optic modulators which is equivalent to modulating the relative phases with respect to beam 1, $\varphi_2$ and $\varphi_3$. This allows for moving the lattice in real space on nearly arbitrary trajectories. In Fig.~\ref{figS1}, our tunable lattice setup is sketched.

The lattice depths $V_l$ stated in the main text are specified in units of the recoil energy $E_r = \hbar^2 k_L^2 / 2m$ with $k_L = 2\pi/\lambda$ and the lattice laser wavelength is $\lambda=1064$~nm, $E_r/h \approx 4410$~Hz. The complete lattice potential can be written as $V(\mathbf{r}) = V_p(\mathbf{r}) + V_s(\mathbf{r})$ with $V_p(\mathbf{r})=V_l \cos^2 (\theta) ( \cos(\mathbf{b}_1\cdot \mathbf{r} - \varphi_2) + \cos(\mathbf{b}_2 \cdot \mathbf{r} + \varphi_2-\varphi_3) + \cos((\mathbf{b}_1+\mathbf{b_2})\cdot \mathbf{r}-\varphi_3) )$ and $V_s(\mathbf{r})=-2V_l \sin^2 (\theta) ( \cos(\mathbf{b}_1)\cdot \mathbf{r} + \alpha_1 - \alpha_2 - \varphi_2) + \cos(\mathbf{b}_2 \cdot \mathbf{r} + \alpha_2 - \alpha_3 + \varphi_2-\varphi_3) + \cos((\mathbf{b}_1+\mathbf{b_2})\cdot \mathbf{r} + \alpha_1 - \alpha_3-\varphi_3) )$. Here, $\mathbf{b}_1 = \mathbf{k}_1 - \mathbf{k}_2$ and $\mathbf{b}_2 = \mathbf{k}_2 - \mathbf{k}_3$ are the reciprocal lattice vectors defined via the wave vectors $\mathbf{k}_1 = k_L (0,1,0)$, $\mathbf{k}_{2/3} = k_L (\pm \sqrt{3}/2,-1/2,0)$. For the experiments presented in the main text we choose $V_l = 15.15(15)~E_r$, $\theta=\pi/20$ and $\alpha_j = (j-1)\cdot 2\pi/3$. In Fig. S1, the resulting dispersion relation of the six lowest bands for the experimental parameters is plotted (lower right corner). Because the bandgap ($11680$~Hz) is large compared to the bandwidths of the two lowest bands ($335$~Hz) and ($48$~Hz), the bands can be considered flat. The lowest two bands are energetically well separated from the other bands. The lattice depth is obtained from the fitted oscillation frequencies $\nu_\mathbf{k}$ (Eq. 1 of the main text) and the dominating systematic error of $1\%$ stems from the lattice depth inhomogeneity across the sample. In the transverse direction perpendicular to the plane spanned by the lattice beams, the potential is harmonic, thus forming an overall lattice of tubes. In addition to the lattice potential, a crossed dipole trap operated at a wavelength of $825$~nm is superimposed, formed by a round beam perpendicular to the lattice direction with a $1/e^2$-intensity beam radius of $w_{x,y} = (70,240)~\mu$m and an elliptical beam propagating in the x-direction with $w_{y,z} = (70,240)~\mu$m. The measured trapping frequencies for the experimental parameters are $\nu_{x,y,z} = (83(4),108(5),93(4))$~Hz. 

\section{Tight-binding description of the lattice Hamiltonian}
\begin{figure}[h]
  \centering
  \includegraphics[width=\columnwidth]{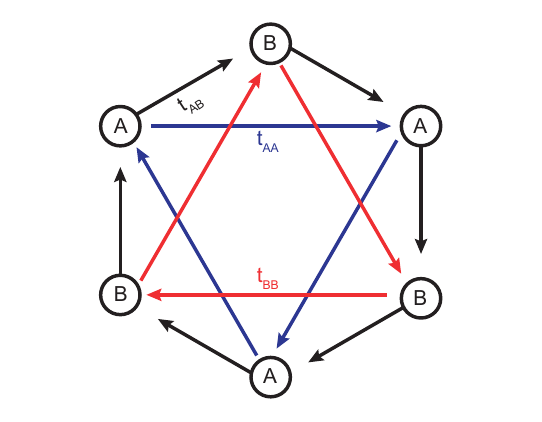}
  \caption{\textbf{Sketch of lattice geometry.} Arrows indicate hopping possibilities $t_{AA}$, $t_{BB}$ and $t_{AB}$. }
  \label{figS2}
\end{figure}

Since the two lowest bands are well separated from the other bands (see Fig.~\ref{figS1}), a two-band tight-binding model~\cite{Ibanez2013} can be employed to describe our system. The tight-binding Hamiltonian can be written as
\begin{widetext}
    \begin{align}
    	\hat{\mathbf{H}} (\mathbf{k}) &= 
		\label{SE1}
    		\begin{pmatrix}
    			\frac{\Delta_{AB}}{2} + \sum_i 2 t_{AA} \cos(\mathbf{k}\cdot \mathbf{a_i}) & \sum_i 2 t_{AB} \exp(- i \mathbf{k}\cdot \mathbf{d_i}) \\
    			\sum_i 2 t_{AB} \exp(+ i \mathbf{k}\cdot \mathbf{d_i}) & -\frac{\Delta_{AB}}{2} + \sum_i 2 t_{BB} \cos(\mathbf{k}\cdot \mathbf{a_i})
    		\end{pmatrix}
    \end{align}
\end{widetext}

with the offset $\Delta_{AB}$ between A- and B-sites and the next-nearest-neighbor hopping amplitudes $t_{AA}$, $t_{BB}$ on the diagonal and on the off-diagonal the nearest-neighbor hopping amplitude $t_{AB}$ coupling the A- and B-sites (see Fig.~\ref{figS2}). 

The vectors $\mathbf{a}_i$ connect neighboring A(B)-sites with each other while the vectors $\mathbf{d}_i$ connect A- with neighboring B-sites. For our perfectly symmetric hexagonal lattice, $\mathbf{a}_1 = -4\pi/3 k_L(0,1,0)$, $\mathbf{a}_1 = -4\pi/3 k_L(\sqrt{3}/2,1/2,0)$ and $\mathbf{a}_3 = \mathbf{a}_2 - \mathbf{a}_1$ are the Bravais vectors, $\mathbf{d}_1 = 1/3(\mathbf{a}_1 + \mathbf{a}_2)$, $\mathbf{d}_2 = 1/3 (-2\mathbf{a}_1 + \mathbf{a}_2)$ and $\mathbf{d}_3 = 1/3(\mathbf{a}_1 - 2\mathbf{a}_2)$. We obtain the tight-binding parameters $\Delta_{AB} = 11.45$~kHz, $t_{AB} = 657$~Hz, $t_{AA} = 32$~Hz and $t_{BB} = 0$~Hz from a fit to the exact band structure shown in Fig.~\ref{figS1}. The asymmetry between the lower and upper band is captured by the differing next-nearest-neighbor hopping amplitudes $t_{AA} \neq t_{BB}$. The uncertainty of $\Delta_{AB}$ stated in the main text arises due to the systematic error of $1~\%$ in the lattice depth (see the previous section). The lower band eigenvectors $\lvert\mathbf{k}\rangle = \sin (\theta_{\mathbf{k}}/2) \lvert\mathbf{k},A\rangle - \cos (\theta_{\mathbf{k}}/2) \exp (i \phi_{\mathbf{k}}) \lvert\mathbf{k},B\rangle$ can be expressed in terms of the angles $\theta_\mathbf{k}$ and $\phi_\mathbf{k}$ which are calculated according to: 
\begin{widetext}
    \begin{align}
    	\sin(\theta) &= \frac{| \sum_i t_{AB}\exp(i\mathbf{k}\cdot\mathbf{d}_i)| }{R}, \cos(\theta) = \frac{\Delta_{AB}/2 + \sum_i (t_{AA}-t_{BB})\cos(\mathbf{k}\cdot\mathbf{a}_i) }{R} \nonumber \\
	\label{SE2}
	\phi &= -\arg \left(\sum_i t_{AB} \exp(i \mathbf{k}\cdot\mathbf{d}_i) \right) \\
	R &= \left(\left(\Delta_{AB}/2 + \sum_i (t_{AA} - t_{BB}) \cos(\mathbf{k}\cdot\mathbf{a}_i)\right)^2 + \left(\sum_i t_{AB} \sin(\mathbf{k}\cdot\mathbf{d}_i)\right)^2 + \left(\sum_i t_{AB} \cos(\mathbf{k}\cdot \mathbf{d}_i)\right)^2\right)^{1/2}. \nonumber
    \end{align}
\end{widetext}

\section{Floquet description of the dressed system}

A periodically driven system can in Floquet theory be described by an effective (time-independent) Hamiltonian $\hat{\mathbf{H}}_{\text{eff}}$ whose properties characterize the dressed system. The effective Hamiltonian is obtained from the unitary time-evolution operator $\hat{\mathbf{U}}(T,0)$ over one period $T$ of the driving~\cite{Kitagawa2010,Hemmerich2010}, $\frac{i h}{T} \log(\hat{\mathbf{U}}(T,0)) = \hat{\mathbf{H}}_{\text{eff}}$. The time evolution operator is given by $\hat{\mathbf{U}}(T,0) = \exp\left(-\frac{i}{h}\int_0^T \hat{\mathbf{H}}(t)dt\right)$ and can be numerically obtained by replacing the integral with the time-ordered product $\hat{\mathbf{U}}(T,0) \approx \prod_{n=0}^{n+1=T/\delta t} \hat{\mathbf{U}}(T-n\Delta t, T-(n+1)\Delta t)$. Using the approximation $\hat{\mathbf{U}}(t+\Delta t,t) \approx \exp\left( -\frac{i}{h} \hat{\mathbf{H}}(t)\cdot \Delta t \right)$, the product can be calculated. We ensured convergence numerically by checking that increasing the amount of time steps does not change the results. Using the tight-binding description of the lattice Hamiltonian as described above (Eq.~\eqref{SE1}) allows for a computationally cost-effective calculation of the effective, time-independent Hamiltonian whose eigenvalues describe the dispersion relation of the dressed bands. Using Eq.~\eqref{SE2}, the angles describing the dressed (effective) eigenvectors on the Bloch sphere are obtained, which allows for a direct comparison with the experimental data as presented in Fig. 3 in the main text.

\section{Derivation of the data for the calculation of the Berry curvature}
To reconstruct the Berry curvature, we fit $n(\mathbf{k},t) = f(\mathbf{k}) \left[ 1-A\exp(-t/\tau) \cos(\phi_{\mathbf{k}} + 2\pi \nu_{\mathbf{k}}t) \right]$ to each pixel, according to Eq. (1) in the main text. The fit parameters are the Wannier envelope $f(\mathbf{k})$, the relative amplitude $A$, the damping constant $\tau$, the phase $\phi_{\mathbf{k}}$ and the frequency $\nu_\mathbf{k}$ which is given by the energy difference between the lower and upper undressed bands. The damping time typically is larger than five oscillation periods. We scale the experimentally measured amplitude by a factor of 1.15, such that the maximum amplitude is 0.92, as expected from theory. We attribute this scaling factor to two effects. First, both the harmonic trap coupling different quasimomenta and the inhomogeneous lattice depth lead to a dephasing and hence reduction of the amplitude of the oscillations. Second, due to unavoidable multi-photon excitations into higher bands which lead to particle loss during the ramp, we choose the ramping time to be not perfectly adiabatic. We quantified the latter by ramping back into the undressed system and a subsequent band mapping and found a relative population in higher bands of $13~\%$. This explains why the experimental amplitude is approximately $15~\%$ smaller than expected from theory and justifies the rescaling. While the phase $\phi_\mathbf{k}$ directly corresponds to the desired tight-binding parameter, the other parameter is obtained via $\theta_\mathbf{k} = \arcsin(A)$. When both the north and south hemispheres of the Bloch sphere are covered (as for non-zero Chern numbers), one can obtain the sign from an additional band mapping measurement or by a careful analysis of the zeros in the amplitude~\cite{Hauke2014}. Before fitting, we smooth $n(\mathbf{k},t)$ along the time direction with a moving average of span 3. For the calculation of the Berry curvature according to Eq. (2), both $\theta_\mathbf{k}$ and $\phi_\mathbf{k}$ and their derivatives are needed. While $\theta_\mathbf{k}$ and $\phi_\mathbf{k}$ are evaluated pixel-wise, their derivatives are obtained from Savitzky-Golay filtered (3$^{\text{rd}}$ order polynomial and window size 15 pixels) finite differences. The pixel-size is small compared to the length ($|\mathbf{b}| = 57(2)$~px) of the reciprocal lattice vectors, allowing for a very high resolution in momentum space. As shown in Fig. 3 in the main text, the experimentally reconstructed Berry curvature is in very good quantitative agreement with the theoretical calculation. As stated in the main text, the Berry phase and Chern number can be directly obtained from the Berry curvature. The Chern number which is given by the integral of the Berry curvature over the first Brillouin zone is in our case simply the sum of Berry curvature over all pixels within the first Brillouin zone (multiplied by the area of the Brillouin zone and divided by $2\pi$). The errors stated for the Chern numbers result from the uncertainty in the length of the lattice vector and an uncertainty of $2$~px in the position of the first Brillouin zone. Since the Berry phase around a loop is the integral of the Berry curvature over a closed area, it simply is the sum of Berry curvature over the pixels in that area multiplied by the area. Hence, the Berry phase can be directly obtained from our data for arbitrary loops, proving the powerfulness of our method.

\cleardoublepage

\end{document}